# Software Defect Prediction Based On Deep Learning Models: Performance Study

Ahmad Hasanpour, Pourya Farzi, Ali Tehrani, Reza Akbari

*Software Engineering Lab, Department of Computer Engineering, Shiraz University of Technology, Shiraz, Iran*
{a.hasanpour, p.farzi, a.tehrani, akbari}@sutech.ac.ir

*Abstract*— in recent years, defect prediction, one of the major software engineering problems, has been in the focus of researchers since it has a pivotal role in estimating software errors and faulty modules. Researchers with the goal of improving prediction accuracy have developed many models for software defect prediction. However, there are a number of critical conditions and theoretical problems in order to achieve better results. In this paper, two deep learning models, Stack Sparse Auto-Encoder (SSAE) and Deep Belief Network (DBN), are deployed to classify NASA datasets, which are unbalanced and have insufficient samples. According to the conducted experiment, the accuracy for the datasets with sufficient samples is enhanced and beside this SSAE model gains better results in comparison to DBN model in the majority of evaluation metrics.

## I. INTRODUCTION

Software quality can be measured by fault proneness data. Some of the newest approaches tried to investigate that whether available metrics in requirement and code could be used to identify fault prone modules. It should be noted that, these metrics and requirement data have been collected during software development cycle and considerable efforts have been deployed to build more accurate defect prediction models by these data to estimate the quality of targeted program modules. In this regard, different approaches have been proposed to predict faulty modules in recent years, like statistical approaches, data mining or machine learning approaches. However, defect prediction models could be applied in different phases in the following classes:

- The first class which is in testing phase include the following models: capture-recapture models [1], neural network models [2], measure method based on scalable method based on source code complexity [3].

- Next class, which was employed to predict number of defects in the software development procedure, is before the actual developing phase of the targeted software. The following models are included in this category: phase based method that is suggested in [4], An Ada-based defect prediction method is proposed in [5], and to predict defections at first stages of programming, a model has been proposed by Smits [6].

The two main problems, which often result in defected data, are high dimensionality and imbalanced classes. In [7], a single classifier approach is presented by Kehan et al., which is based on data sampling and feature selection to deal with the aforementioned problems. They consider three scenarios, such that feature selection is based on two different types of data, i.e., original data or sampled data. They concluded that the scenario which is feature selection have done on sampled data and have modeled on original data have significantly higher performance than the other scenarios. Additionally the performance could be boosted by using ensemble classifiers instead of single classifier.

Software defect prediction using ensemble learning has been investigated by Tao WANG et al. [8]. They presented comprehensive results on the ensemble learning methods that voting and random forest achieve higher performance than other classifiers. Moreover, Meta ensembling is shown to have better generalization ability in WANG method.

Despite all the recent research, ensemble defect prediction does not turn in great performance time after time due to the diversity between the data distributions for training and testing part in ensemble approaches. It is clear that majority of classifiers give a high performance when train data have as same feature and space as test data do. Furthermore data distribution should be the same to achieve this goal [9] but defect data usually have different distribution and feature space. Deep learning models cover some drawbacks of machine learning algorithms because of able to learn through its own computing brain. Recently they have become more popular as far as different types of applications ranging from pattern analysis and classification have been developed by researches using these techniques. Therefore, we decided to deploy these models for defect problems.

Beside the above mentioned problems, it is difficult to achieve a reasonable defect prediction when considering a large software system, because selecting and testing all software modules would take a great amount of time. Therefore, in [10] sample-based methods are provided to tackle this issue for defect prediction in software. Three categories were considered for these methods: active sampling with active semi-supervised-based classifiers, random sampling with a semi-supervised-based classifiers, and random sampling with conventional-based classifiers. However, it should be mentioned that sampling approaches decrease accuracy of the defect prediction. Deep learning techniques could be better confront with accuracy rate especially when the number of samples are many. The following paragraph is a brief introduction of deep learning.

Deep learning is known as a sub-category of machine learning techniques. In this type of learning new structures are used where many layers of processing units are employed for feature extraction and transformation. Similar to traditional neural networks, each layer uses the output from the previous layer as its input. In addition, supervised and unsupervised learning can be assumed. Moreover, it is deployed as a new solution for most of the domains especially in software topics. Commercial applications that use open source platforms with consumer recommendation programs [11], image recognition [12] and medical research tools [13] that explore the possibility of reusing drugs for new ailments are a few of the deep learning applications examples. In this paper, the applicability of two deep learning methods is studied for software defect prediction problem. We employed two generative deep learning models as they are DBN and SSAE. For evaluation and testing phase, PROMISE datasets are employed. The obtained results show that our technique give a magnificent performance for software defect prediction.

This research is consisted of seven part that are arranged carefully as follows: in section II, some of the most prominent and recent previous works will have been discussed and explain. The details of deep learning methods are described in Section III. In addition, test settings and obtained results are discussed in Sections IV and V respectively. Performance measurements and final analysis are given in Section VI. Finally, a firm conclusion had been written in Section VII.

## II. RELATED WORKS

In recent years, wide variety of deep learning models have been proposed and applied to different domains by researchers. However, in defect prediction context, to the best of our knowledge few works have been done which we will review in this section.

*TABLE 1. A taxonomy of the related works along with the proposed method*

| Approach | PAPER | Method (Mining, Learning, Optimization) | Methodology | Preprocessing step | Class imbalance problem | Supervised Semi-supervised | Datasets |
|---|---|---|---|---|---|---|---|
| The Proposed Method | - | Deep Learning | DBN, SSAE | Normalization | Not Considered | - | **CM1, KC1, KC2, KC3, KC4, PC1, PC2, PC3, PC4, PC5, JM1, MW1, MC1, MC2** |
| WPDP | [17] | Minining & Learning | OneR, J48, and naïve Bayes | removing the module identifier attribute | Not Considered | Supervised | KC3,CM1,KC4,MW1,PC1, PC2,PC3,PC4 |
| WPDP | [18] | Minining & Learning | Extended transfer component analysis +logistic regression | min-max and z-score normalization methods | Not Considered | Supervised | ReLink,AEEEM |
| WPDP | [19] | Minining & Learning | WC and CC-data models | NN-filtering | Not Considered | Supervised | PC1,KC1,KC2,CM1,KC3, AR3,AR4,AR5 ,MW1,MC2 |
| CPDP | [20] | Learning | Transfer Naive Bayes | NN-filtering | Not Considered | Supervised | kc3,Pc1,kc1,kc2, cm1, ar3,ar4,ar5, mw1,mc2 |
| CPDP | [21] | Mining & Learning | context-aware rank transformations | Clean Data (Understand) | Not Considered | Semi-supervised | Generate a dataset |
| CPDP | [22] | Learning | ensemble approaches | Minimizing collinearity | Considered | Supervised | Bugzilla ,Columba ,Gimp ,Eclipse JDT ,Maven-2 ,Mozilla |
| HDP | [23] | Learning | canonical correlation analysis (CCA) nearest neighbor (NN) | z-score normalization | Considered | Supervised | NASA,SOFTLAB,ReLink AEEEM |
| HDP | [24] | Learning | Logistic regression | Feature selection (gain ratio, chi-square, relief-F) | Considered | Supervised | AEEEM,ReLink,MORPH NASA,SOFTLAB |

Y. Chen and et al [14] reviewed the previous work in field of defect management and software prediction. They introduce a novel method for defect prediction by using data mining techniques and claim that their proposed model is able to lead the developmental stages of a new software. At first, defect database is generated which is made up of all of the information about the defect data in the software life cycle. After that, by mining techniques, in particular Bayesian Network, the defect prediction model is constructed for the going.

An enhanced multilayer perceptron neural network is explored by [15], and also fault-proneness prediction modeling is performed by comparative analysis for software systems and then tested by NASA's Metrics Data Program (MDP). Gabriela Czibula et al. present in [16] a novel classification model regarding relational association rules mining.

Identifying defective modules is not always a straightforward task. To attain high performance, various aspect should be considered in defect prediction models. Ishani and Arora and et al. in [25] introduce some of them in detail. Their research show that these issues are caused by the following problems:

- Relationship between Attributes and fault.
- No benchmark to assess performance correctly.
- Issues with defect prediction in cross-project.
- No available general framework.
- Economic limitations of defect prediction in software.
- Class imbalance issue.

Also [13] noted wide variety of models for example bagging and boosting [26], [27] and Naïve-Bayes [17], [27], [28], one rule [17], [28], [29], SVM [30]–[32], J48 decision tree [17], [28], [31] and etc.

Besides the above methods, association based classification approach is considered in this context by Baojun Ma and et al [33]. They use CBA2 algorithm and compare it with the other rule based classification methods. Their experimental results shows CBA2 acts better than C4.5 and RIPPER algorithms.

Generally, defect prediction procedure is designed by supervised machine learning (classification), which is called within-project defect prediction (WPDP) because all processes are conducted 'within' a single software project. Some preprocessing techniques such as feature selection and normalization are broadly applied in these studies [17], [18], [19]. However, WPDP has some intrinsic limitations since training models without information of defect data generate the labeled dataset. Researchers have also proposed techniques to improve cross-project defect prediction (CPDP) [18]–[22], [34], [35] that is defect prediction for unlabeled datasets [36], [37]. CPDP generally has low performance. However, most CPDP approaches have some limitations that have some significant effects on the performance for instance; they should use same metrics if source data set and target data set had heterogeneous metrics. In order to solve this problem Jaechang Nam and Sunghun Kim [24] presented a new algorithm. They proposed heterogeneous defect prediction (HDP) method for predicting defects across project sets (whether heterogeneous metrics exist in dataset). Indeed, the source project and destination project can be different from each other. We classified most important recent research based on three categories as they are considered in TABLE 1. Six parameters are discussed in details for methods in TABLE 1. In total, the following results are observed from TABLE 1 (Taxonomy table) by scrutiny of the models:

- The learning methods are deployed for sample classifications in most approaches.
- The class imbalanced problem was not taken in to account in most recent research and they try to robust their approaches.
- The majority of recent approaches have considered NASA's data set as their base; therefore, our experiments are also on this dataset.
- A vast number of approaches did not consider preprocessing step, while a few of others have complicated method for doing it.

This paper presents a new way to improve defect prediction by leveraging the power of two deep learning models. The steps are discussed in detail after introducing deep learning models.

### III. DEEP LEARNING MODELS

The proposed scheme is shown in the Fig.1. Following that, each step is discussed in detail. The scheme is designed based on our experiment, which are consist of four steps as it is shown in Fig.1.

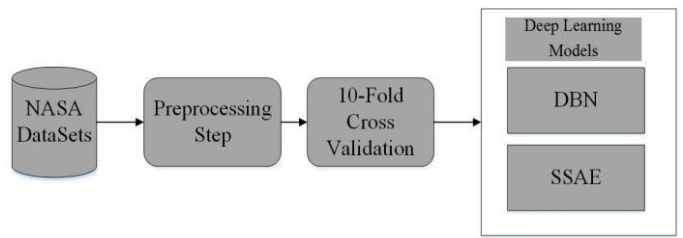

Fig.1. the general view of the proposed model

As it was mentioned earlier, two neural networks are used in order to conduct this research. The first network is DBN, which is based on RBM (Restricted Boltzman Machine) and composed of multiple layers. The second network is SSAE that is the extension of auto-encoder and is made up of multi-layer sparse auto-encoders. A DBN is capable of learning to reconstruct its input with a degree of probabilistic. Each layer in DBN then represent as detector for input features. When the learning process is done, it can be trained even further in a supervised style to conduct classification. In this type of networks, a sparseness constraint is applied on hidden units and then data structure will be discovered considering the large number of hidden units.

## A. Deep Belief Networks

G.Hinton [38] introduced this network to overcome the major limitation (for example these models were generally restricted to only a few layers) of earlier neural network. DBN is made of two types of neural networks (Belief network and restricted Boltzmann machine) and is a type of unsupervised algorithms. DBN is a generative graphical model build on RBM. It learns a representation from input data, and an output will be constructed from a representation that contains contents and semantics of input data.

In general, DBN has three main parts [39], the lowest part is input layer, and the next two parts are multiple hidden layers and output layer, respectively. The main notion of DBN, which usually uses restricted Boltzmann machines as basic blocks, is to give the reconstruction ability to the network thus making it able to renovate input data by means of adjusting the weights between nodes in different layers. Beside this, it has two training phases include RBM pretraining at each layer and fine tuning of the stacked RBMs. Restricted Boltzman Machine is an undirected, bipartite graphical model, which composes of two layers the first of which includes data variables, which are called visible nodes, while the second layer includes hidden variable that are referred as hidden nodes. As it is shown in Fig.2, RBM topology scheme is a completely connected by way of undirected weight edges which are symmetric.

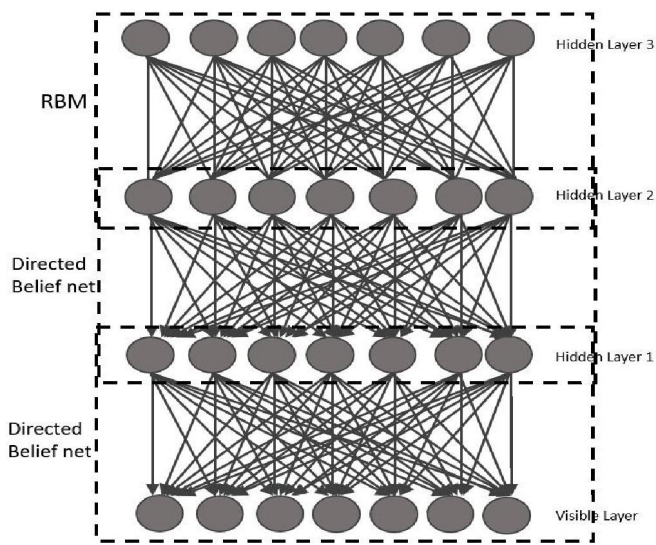

Fig.2. DBN Scheme

A DBN can be built by stacking the RBMs and adding a softmax regression layer. In this network, we need to model the joint distribution which connect hidden and input layers, all of which are trained as an RBM according to the formula that is described in details in [39]. Hidden layer's state in the lowest level of RBM is considered as the input source to the upper RBM (visible layer). Moreover, state vector, which obtained from the uppermost hidden layer, is applied as an input to softmax regression layer.

## B. Stack Sparse Auto-Encoder

Stack auto-encoder, which is classified as a learning algorithm, is composed of a stack of auto-encoders. Data correlation discoverer focuses on representation of different features that are obtained from input data. It should be mentioned that aforementioned data are high dimensional. Auto-encoder is a type of neural network in which there are multi-layer that learned to encode the input data with back-propagation, by which discrepancy among inputs and reconstruction will be decreased.

The main idea behind an auto-encoder is that, an auto-encoder should create y in such way that represents the main structures in input. Imagine that an auto-encoder has N hidden units and its input has M units, where $N < M$, in this case, auto-encoder should compress and present the input in way that can have the ability to be reconstructed efficiently.

The architecture of SSAE is indicated in Fig.3. The input section is comprised of an encoder that changes input x to the corresponding delineation h, which can be considered as a new picture of input data. Furthermore, the output section represent a decoder that is trained to construct an estimation of input from the hidden delineation.

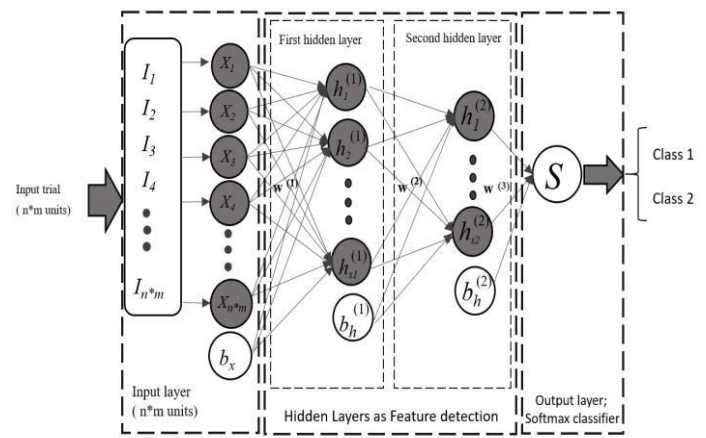

Fig.3. SSAE Scheme

A serious problem that most auto-encoders have to deal with is that sometimes they cannot extract useful features from input, this problem gets severe specially when hidden units have as many number as input units or in some cases have greater than input units. To solve this problem, it is common that sparsity constraints, which are discussed in [40], are applied on auto-encoders. This type of auto-encoders is usually named Sparse Auto-encoders.

In order to create deep network with Stacked Auto-encoders, Stacked Auto-encoder uses the explained auto-encoders as building blocks. If sparse auto-encoders are used then it is called Stacked Spare Auto-encoder (SSAE).

## IV. EXPERIMENTAL DESIGN

### A. Datasets

The public NASA datasets, which consist of information related to several NASA projects, deployed in this research for evaluation. In recent years, these datasets have drawn a great amount of attentions from researchers in this field. The statistics of the datasets are shown in the TABLE 2.

Name of fourteen dataset have been shown in TABLE 2. Information about the programming language and Lines of Code (LOC) and the percentage of defective modules for each dataset is shown in other three columns respectively.

TABLE 2. The detailed information of NASA datasets

| Data set | Language | LOC | # sample (positive,negative) |
|---|---|---|---|
| CM1 | C | 20k | 505(0.095,0.905) |
| KC1 | C++ | 43k | 2107(0.154,0.845) |
| KC2 | Java | 18k | 522(0.201,0.798) |
| KC3 | Java | 18k | 458(0.093,0.906) |
| KC4 | Perl | 25k | 125(0.6,0.4) |
| MC1 | C++ | 63k | 9466(0.007,0.992) |
| MC2 | C | 6k | 161(0.322,0.677) |
| PC1 | C | 40k | 1107(0.068,0.931) |
| PC2 | C | 26k | 5589(0.004,0.995) |
| PC3 | C | 40k | 1563(0.102,0.898) |
| PC4 | C | 36k | 1458(0.122,0.878) |
| PC5 | C++ | 16k | 17186(0.030,0.970) |
| JM1 | C | 315k | 10878(0.19,0.81) |
| MW1 | C | 8k | 403(0.08,0.92) |

*B. Preprocessing Step (normalization)*

Since each sample contains different values that can vary greatly, feature scaling, which is one of popular methods in normalization, is performed to normalize features (independent variables). To achieve this goal, standardization method is selected and used for this section. Standardization is widely used for normalization in many machine learning algorithms. Feature normalization is done according to the formula below,

$$\frac{D_i - \mu_i}{\sigma_i} \quad (1)$$

Where $D_i$ defines the ith data dimension, $\mu_i$ is average, and $\sigma_i$ defines standard deviation of that dimension.

*C. K-Fold Cross Validation*

After normalization step, a 10-fold cross-validation strategy is applied to compute the parameters of the test set. Each dataset is randomly partitioned into 10 subsets, each of which is equal to others in terms of its size and one of which is considered as test data every time while the other nine subsets are considered as training data. This action should be reiterated ten times for running same algorithm on data. Finally, the mean of these ten runs is computed.

## V. EXPERIMENTAL RESULTS

Each deep learning model has a variety of parameters such as number of epochs and batch size that need to be set. Finding and assigning best optimal value to each of these parameters is a challenging task. This is usually done via training our model several times with different values assigned to parameters. Doing this, will enable us to choose the optimum-observed values for each parameter. TABLE 3 shows the final values of the parameters for each deep learning model.

TABLE 3. Fixed parameters in each model.

| Models | Parameters |
|---|---|
| DBN | pre-training epoch=20, batch size=4, epochs=150, fine tune learning rate=0.01, pretrain learning rate=0.001 |
| SSAE | pre-training epoch=50, batch size=4, fine tune learning rate=0.01, epochs=150, $\rho = 0.05$ |

On the other hand, a number of parameters should be tested several times for each dataset such as the number of middle layers and neurons. The values in TABLE 4 are obtained from the optimal configurations. The [X,Y] notation specifies two layers where X and Y are the numbers of neurons in first and second layer, respectively. Increasing the number of layers and neurons is possible to recognize problematic patterns (when system is faced with high volume and dimensions of data).

TABLE 4. The optimal parameters for datasets

| Dataset | Parameters |
|---|---|
| CM1 | DBN [30,12] |
|  | SSAE [25,15,7] |
| KC1 | DBN [20,15,10] |
|  | SSAE [25,15,8,4] |
| KC2 | DBN [20,10] |
|  | SSAE [20,10] |
| KC3 | DBN [15,5] |
|  | SSAE [15,10] |
| KC4 | DBN[15,5] |
|  | SSAE[15,8] |
| MC1 | DBN [40,25,10] |
|  | SSAE [40,30,20,10] |
| MC2 | DBN [30,10] |
|  | SSAE [30,15] |
| PC1 | DBN [20,15,10] |
|  | SSAE [20,10,10,5] |
| PC2 | DBN [20,10,10,10] |
|  | SSAE [20,20,10,10,10] |
| PC3 | DBN [20,10] |
|  | SSAE [20,10,10] |
| PC4 | DBN [30,20,10] |
|  | SSAE [25,20,10,10] |
| PC5 | DBN [35,30,20,10,8] |
|  | SSAE [35,30,20,20,10] |
| JM1 | DBN [50,30,20,8] |
|  | SSAE [40,30,10,8] |
| MW1 | DBN [30,15,4] |
|  | SSAE [30,15,4] |

Our experiment results are listed and summarized in TABLE 5 (in percentages) and compared with VOTE classifier that is reported in [8] as one of the best mean of accuracy on all datasets. To show the efficiency of proposed approach, in addition to VOTE, the results are also compared against other prominent methods used in the literature such as CSVS + CSNN [41], CSLS + CSNN [41], CBA2 [33] and SVM + static code metrics [42]. The other evaluation parameters respect to the different criteria such as recall, miss rate, etc, are shown and comprised with the other methods in TABLE 6. As it can be seen from the obtained results, there is a difference between SSAE and DBN with the VOTE method for NASA datasets.

TABLE 4 reports the percentage of accuracy obtained from our experiment. It also compares the results with other prominent methods used in literature, in particular, VOTE [8], CSVS + CSNN [40], CSLS + CSNN [40], CBA2 [32] and SVM with static code metrics [41]. Bellow the percentage in each cell is the rank of that corresponding method comparing with others in the same row.

As it was mentioned earlier, one of the issues that needs to be taken care of is data unbalancing. This issue also affects the error correction in some cases where there are not sufficient positive samples. The deep learning models used in this research are generative, that means they can generate new samples. In these models, each layer is pertained to input data distribution and is independent from one another. Generating new samples similar to other samples exist in datasets helps to reduce the impact of data unbalancing. Despite using generative models, in some cases where the number of samples is insufficient the models still show low accuracy.

TABLE 5. The experimental results of accuracy parameter comparison based on deep learning approach

| Dataset | DBN | SSAE | VOTE | CSVS + CSNN | CSLS + CSNN | CBA2 | SVM |
|---|---|---|---|---|---|---|---|
| CM1 | 88.57 ±1.9 | 88.59 ±2.61 | **89.64 ± 2.30** | 77.60 ± 0.42 | 74.44 ± 0.56 | 80.36 | 68 |
| KC1 | 85.83 ±0.86 | **85.63 ±1.23** | 85.62 ± 1.64 | - | - | 83.71 | - |
| KC2 | 81.60 ±1.1 | **84.48 ±0.85** | 82.91 ± 3.38 | 74.07 ± 0.59 | 74.82 ± 0.68 | - | - |
| KC3 | 75.36 ± 0.52 | 77.60 ±2.8 | **89.98 ±3.20** | - | - | 90.91 | 66 |
| KC4 | 69.59 ±0.8 | 69.60 ±1.6 | 75.38 ±11.43 | - | - | **85.37** | 71 |
| PC1 | 92.51 ±0.78 | **94.13 ±1.46** | 93.73 ± 1.45 | 83.73 ± 1.93 | 82.01 ± 2.23 | 91.78 | 71 |
| PC2 | 97.79 ±0.11 | 99.39 ±0.08 | 99.53 ± 0.13 | **99.63 ± 0.11** | 99.19 ± 0.20 | 99.20 | 64 |
| PC3 | 87.26 ±0.72 | **90.21 ±0.97** | 89.12 ± 1.77 | 75.80 ± 0.39 | 78.80 ± 0.18 | 86.48 | 76 |
| PC4 | 88.06 ±0.48 | **91.22 ±1.17** | 90.28 ± 1.75 | 82.23 ± 1.09 | 85.00 ± 0.25 | 83.96 | 82 |
| PC5 | 97.07 ±0.66 | **97.47 ±0.74** | 97.46 ± 0.23 | - | - | - | 69 |
| JM1 | 81.32 ±0.12 | **84.59 ±0.65** | 81.44 ± 0.56 | - | - | 73.52 | - |
| MW1 | 92.55 ±0.53 | **93.30 ±1.78** | 91.67 ± 3.07 | 87.93 ± 0.43 | 85.06 ± 0.59 | 91.04 | 71 |
| MC1 | 99.12 ±0.04 | **99.53 ±0.12** | 99.42 ± 0.13 | - | - | 95.00 | 65 |
| MC2 | 59.62 ±3.10 | 61.49 ±4.75 | **72.57 ± 7.14** | - | - | 69.81 | 64 |
| Weighted Rank | 3 | 1 | 2 | 6 | 7 | 4 | 5 |

## VI. PERFORMANCE MEASUREMENT AND DISCUSSION

There are a number of terms, which are used as standard metrics for comparing the results namely performance evaluation parameters, which include of false positive, true negative, true positive, and false negative. These parameters compare the classifiers as a fair measure. Confusion matrix is calculated for each algorithm and for each dataset. Beside these metrics, some other metrics are derived from the matrix to improve the transparency of better evaluation and comparison. These metrics are shown in the TABLE 6.

In addition, Accuracy is one of the prominent metrics for evaluation phase, because this measure determines what proportion of items are correctly classified.

Fig.4. shows the mean rate of error based on the epoch number after training for 10 times. As is seen from the given illustration, the mean error rate of DBN in early epochs are less than that of SSAE model. Moreover, DBN also reaches the convergence points faster, meaning fewer epochs are required.

TABLE 6. The comparison of performance evaluation metrics

| Datasets | Algorithms | Recall | Accuracy | Precision | Positive Likelihood Ratio (LR+) | Negative Likelihood Ratio (LR-) |
|---|---|---|---|---|---|---|
| CM1 | SSAE | 0.97 | 0.90 | 0.92 | 1.22 | 0.16 |
| | DBN | 0.95 | 0.88 | 0.91 | 1.09 | 0.37 |
| KC1 | SSAE | 0.95 | 0.86 | 0.89 | 1.43 | 0.14 |
| | DBN | 0.96 | 0.86 | 0.88 | 1.41 | 0.13 |
| KC2 | SSAE | 0.91 | 0.81 | 0.87 | 1.45 | 0.21 |
| | DBN | 0.9 | 0.83 | 0.87 | 1.8 | 0.16 |
| KC3 | SSAE | 0.82 | 0.77 | 0.92 | 1.13 | 0.66 |
| | DBN | 0.80 | 0.75 | 0.91 | 1.07 | 0.79 |
| KC4 | SSAE | 0.83 | 0.70 | 0.66 | 1.87 | 0.31 |
| | DBN | 0.83 | 0.70 | 0.66 | 1.87 | 0.31 |
| PC1 | SSAE | 0.98 | 0.94 | 0.95 | 1.58 | 0.04 |
| | DBN | 0.99 | 0.93 | 0.94 | 1.1 | 0.13 |
| PC2 | SSAE | 0.99 | 0.99 | 0.99 | 1.04 | 0.05 |
| | DBN | 0.98 | 0.98 | 0.99 | 0.98 | 0 |
| PC3 | SSAE | 0.97 | 0.9 | 0.92 | 1.39 | 0.1 |
| | DBN | 0.95 | 0.87 | 0.92 | 1.23 | 0.23 |
| PC4 | SSAE | 0.99 | 0.91 | 0.91 | 1.4 | 0.04 |
| | DBN | 0.97 | 0.88 | 0.9 | 1.25 | 0.12 |
| PC5 | SSAE | 0.99 | 0.97 | 0.98 | 1.42 | 0.02 |
| | DBN | 0.99 | 0.97 | 0.98 | 1.37 | 0.03 |
| JM1 | SSAE | 0.99 | 0.85 | 0.84 | 1.25 | 0.01 |
| | DBN | 0.99 | 0.81 | 0.82 | 1.08 | 0.15 |
| MW1 | SSAE | 0.99 | 0.93 | 0.93 | 1.15 | 0 |
| | DBN | 0.99 | 0.93 | 0.93 | 1.03 | 0 |
| MC1 | SSAE | 0.99 | 0.99 | 0.99 | 1.58 | 0 |
| | DBN | 0.99 | 0.99 | 0.99 | 1.04 | 0.04 |
| MC2 | SSAE | 0.85 | 0.61 | 0.67 | 0.96 | 1.27 |
| | DBN | 0.83 | 0.6 | 0.66 | 0.92 | 1.72 |

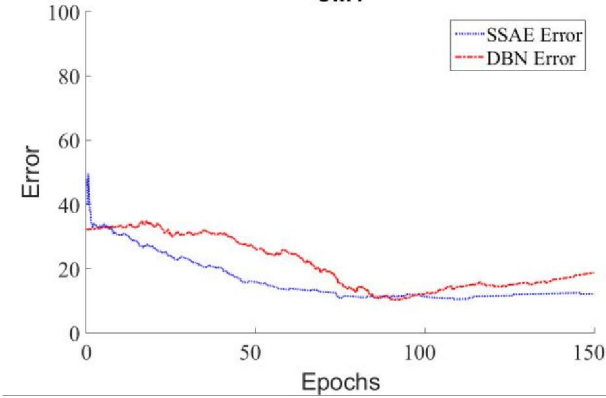
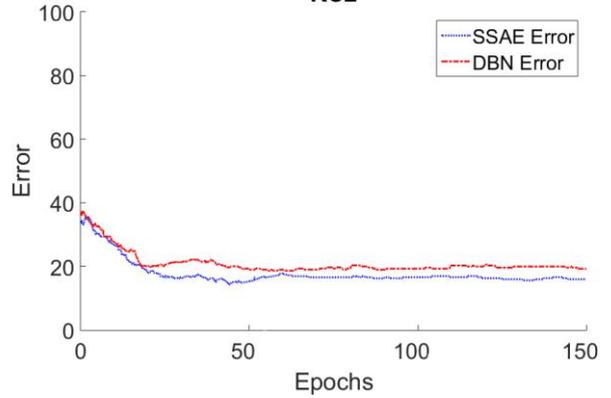
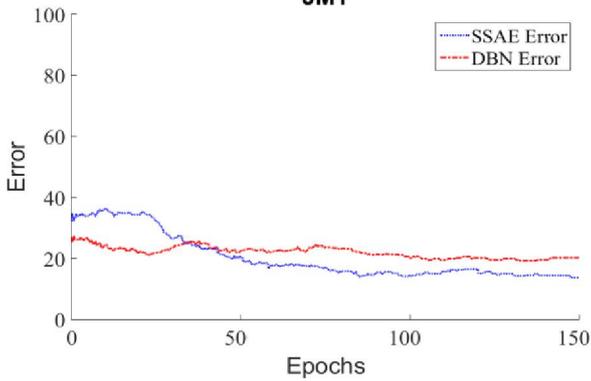
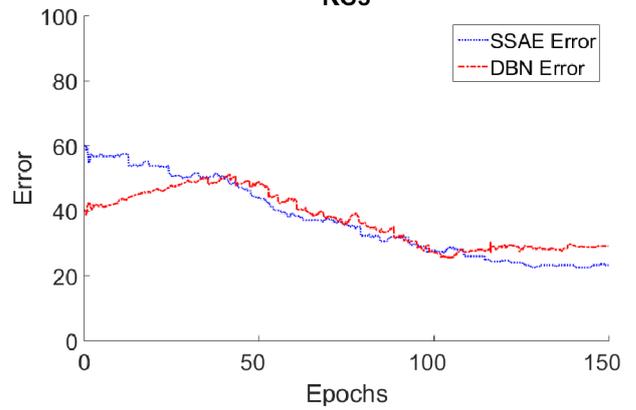
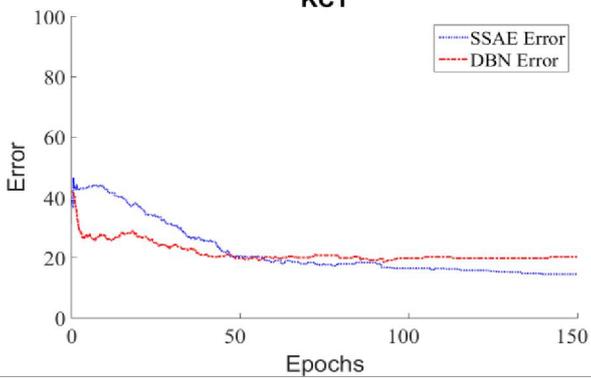
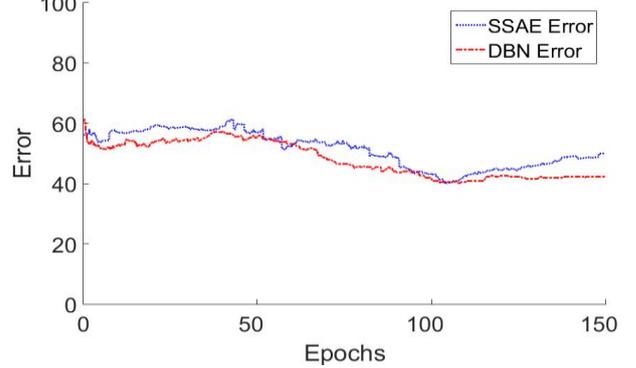

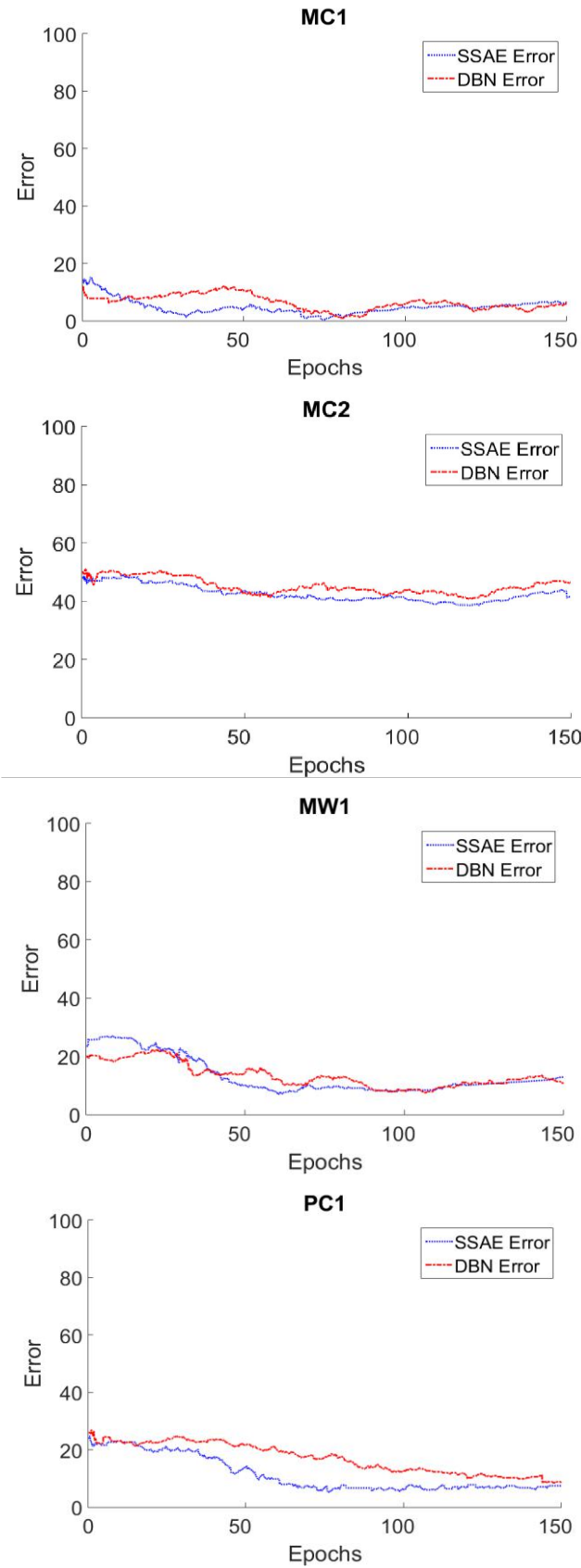
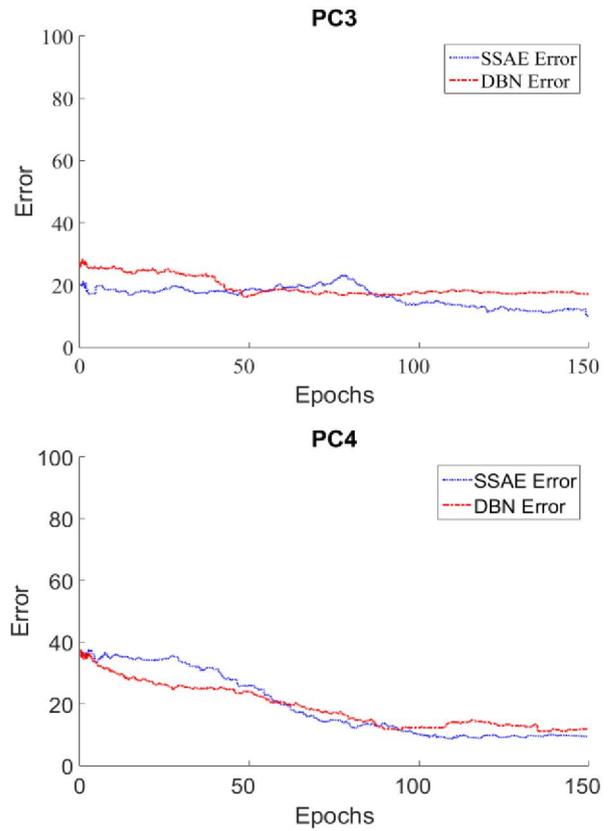

Fig.4. Error diagrams of various datasets classified by SSAE and DBN

Both of the aforementioned problems (data unbalancing and insufficient number of samples) are existed in MC2, KC4, and KC3, and none of the deep learning models could do classification properly for these three datasets. It worth mentioning that precision and recall are also at their lowest level for these three datasets in deep models.

However, it seems that classifiers could improve evaluation metrics in some datasets in spite of having unbalanced data of the existence of unbalancing data. JM1 is a good example, which has sufficient samples and better data balancing. SSAE algorithm shows higher accuracy for JM1 dataset and distribution of two classes are trained based on the deep learning models as it can be seen in the measurement metrics.

## VII. CONCLUSION

We have used generative deep learning models for the first time in defect prediction process. As our experimental results h shows, these models achieve higher accuracy than the other approaches. Although for some datasets the obtained results are impressive, some other datasets, which do not have sufficient samples and suffer from data unbalancing engendered poor results. However, SSAE presents the best generalization ability with accuracy numerical mode. In future, we will investigate how other deep learning models can affect the results. In addition, we plan to apply augmentation models for balancing the data and evaluate their effectiveness.